\documentstyle[referee]{laa}

\begin{document}
\thesaurus{07  
	   (07.09.1; 
	    07.13.1;  
	   )}

\title{Electromagnetic Radiation and Motion of Real Particle}
\author{J.~Kla\v{c}ka}
\institute{Institute of Astronomy,
   Faculty for Mathematics and Physics, Comenius University \\
   Mlynsk\'{a} dolina, 842~48 Bratislava, Slovak Republic}
\date{}
\maketitle

\begin{abstract}
Relativistically covariant equation of motion for real dust particle
under the action of electromagnetic radiation is derived. The particle
is neutral in charge. Equation of motion is expressed in terms of
particle's optical properties, standardly used in optics for stationary
particles.


\end{abstract}

\section{Introduction}
Relativistic equation of motion for perfectly absorbing spherical dust particle
under the action of electromagnetic radiation was derived by
Robertson (1937). Relativistic generalization for the case when scattered
radiation is in the direction of the incident radiation -- in proper reference
frame of the particle -- was presented by Kla\v{c}ka (1992, 2000).

However, real particles scatter radiation in a more complicated manner.
The consequence of this reality may be a completely different orbital
evolution of dust particles. Kocifaj and Kla\v{c}ka (1999) show this
for really shaped stationary rapidly rotating particle.

Equation of motion for real (cosmic) dust particle was presented
in Kla\v{c}ka (1994a)
(and applied to real situation in Kla\v{c}ka 1994b).
Paper by Kla\v{c}ka and Kocifaj (1994) expresses
the equation in terms of optical properties used in optics.
However, the last two papers express the result only to the first order in
$\vec{v}/c$ (higher orders are neglected) where $\vec{v}$ is velocity
of the particle, $c$ is the speed of light. Such an accuracy
may be sufficient in many applications in practice. However,
to be sure that the equation of motion presented in the above two
papers is correct, one has to derive relativistically correct equation
of motion taking into account all orders in $\vec{v}/c$.

The aim of this paper is to derive
relativistically covariant equation of motion for real dust particle
under the action of electromagnetic radiation.

\section{Proper reference frame of the particle -- stationary particle}
The term ``stationary particle'' will denote particle which does
not move in a given inertial frame of reference,
although we admit its rotational
motion around an axis of rotation (with negligible rotational velocity).
Primed quantities will denote quantities measured in the
proper reference frame of the particle.

The flux density
of photons scattered into an elementary solid angle
$d \Omega ' = \sin \vartheta ' ~ d \vartheta ' ~ d \varphi '$
is proportional to  $p' ( \vartheta ', \varphi ') ~ d \Omega '$, where
$p' ( \vartheta ', \varphi ')$ is ``phase function''.
Phase function depends
on orientation of the particle with respect to the direction of the
incident radiation and on the particle characteristics;
angles $\vartheta '$, $\varphi '$ correspond to the direction (and orientation)
of travel of the scattered radiation, $\vartheta '$ is polar angle
and it equals zero for the case of the travel of the ray in the orientation
identical with the unit vector $\hat{\vec{S}_{i}} '$ of the incident radiation.
The phase function fulfills the condition
\begin{equation}\label{1}
\int_{4 \pi} p' ( \vartheta ', \varphi ')~ d \Omega ' = 1 ~.
\end{equation}

The momentum of the incident beam of photons which is lost in the process
of its interaction with the particle is proportional to the cross-section
$C'_{ext}$ (extinction). The part proportional to $C'_{abs}$ (absorption)
is completely lost and the part proportional to
$C'_{ext} ~-~ C'_{abs} = C'_{sca}$ (scattering) is again reemitted.

The momentum (per unit time) of the scattered photons into an elementary
solid angle $d \Omega '$ is
\begin{equation}\label{2}
d \vec{p'}_{sca} = \frac{1}{c} ~ S' ~ C'_{sca} ~
		   p' ( \vartheta ', \varphi ')~ \hat{\vec{K'}} ~ d \Omega ' ~,
\end{equation}
where unit vector in the direction of scattering is
\begin{equation}\label{3}
\hat{\vec{K'}} = \cos \vartheta '~\hat{\vec{S}_{i}} ' ~+~
		 \sin \vartheta ' ~ \cos \varphi ' ~ \hat{\vec{e}_{1}} ' ~+~
		 \sin \vartheta ' ~ \sin \varphi ' ~ \hat{\vec{e}_{2}} ' ~.
\end{equation}
$S'$ is the flux density of radiation energy. The system of unit vectors
used on the RHS of the last equation forms an orthogonal basis.
The total momentum (per unit time) of the scattered photons is
\begin{equation}\label{4}
\vec{p'}_{sca} = \frac{1}{c} ~ S' ~ C'_{sca} ~ \int_{4 \pi} ~
		 p' ( \vartheta ', \varphi ')~ \hat{\vec{K'}} ~ d \Omega ' ~.
\end{equation}

The momentum (per unit time)
obtained by the particle due to the interaction
with radiation is
\begin{equation}\label{5}
\frac{d~ \vec{p'}}{d~ t'} = \frac{1}{c} ~ S' ~ \left \{
			      C'_{ext} ~\hat{\vec{S}_{i}} '
			      ~-~ C'_{sca} ~ \int_{4 \pi} ~
			      p' ( \vartheta ', \varphi ')~ \hat{\vec{K'}} ~
			      d \Omega ' \right \} ~.
\end{equation}

As for the energy, we suppose that it is conserved: the energy (per unit time)
of the incoming radiation $E'_{i}$, equals to the energy (per unit time)
of the outgoing radiation (after interaction with the particle) $E'_{o}$.
We will use the fact that time $t' = \tau$, where $\tau$ is proper
time.

For the sake of brevity, we will use ``effective factors'' $Q'_{xxx}$
instead of effective cross-sections $C'_{xxx}$:
$C'_{xxx} = Q'_{xxx} ~ A'$, where $A'$ is geometrical
cross-section of a sphere of volume equal to the volume of the
particle. Equation (5) can be rewritten to the form
\begin{eqnarray}\label{6}
\frac{d ~\vec{p'}}{d~ \tau} &=& \frac{1}{c} ~ S'~A'~ ~ \left \{ \left [
	     Q'_{ext} ~-~ < \cos \vartheta'> ~ Q'_{sca} \right ] ~
	     \hat{\vec{S}_{i}} ' ~+~ \right .
\nonumber \\
& &  \left . \left [ ~-~ < \sin \vartheta' ~ \cos \varphi ' > ~ Q'_{sca}
	     \right ] ~ \hat{\vec{e}_{1}} ' ~+~
	     \left [ ~-~ < \sin \vartheta' ~ \sin \varphi ' > ~ Q'_{sca}
	     \right ] ~ \hat{\vec{e}_{2}} ' \right \} ~,
\end{eqnarray}
or, in a short form
\begin{equation}\label{7}
\frac{d~ \vec{p'}}{d~ \tau} = \frac{1}{c} ~ S'~A'~ ~ \left \{
	     Q_{R} ' ~ \hat{\vec{S}_{i}} ' ~+~ Q_{1} '
	     ~ \hat{\vec{e}_{1}} ' ~+~ Q_{2} '
	     ~ \hat{\vec{e}_{2}} ' \right \} ~.
\end{equation}

\subsection{Summary of the important equations}
Using the text concerning energy below Eq. (5) and the last Eq. (7), we
may describe the total process of interaction in the form of the
following equations (energies and momenta per unit time):
\begin{eqnarray}\label{8}
E_{o} ' &=& E_{i} ' = A'~S' ~,
\nonumber \\
\vec{p}_{o} ' &=& ( 1 ~-~ Q_{R} ' ) ~ \vec{p}_{i} ' ~-~
		  (  Q_{1} ' ~ \hat{\vec{e}_{1}} ' ~+~
		     Q_{2} ' ~ \hat{\vec{e}_{2}} ' ) ~ E_{o} ' / c ~,
\nonumber \\
\vec{p}_{i} ' &=& ( E_{i} ' / c ) ~ \hat{\vec{S}_{i}} ' ~,
\end{eqnarray}
The index $"i"$
represents the incoming radiation, beam of photons, the index $"o"$
represents the outgoing radiation.

The changes of energy and momentum of the particle due to the interaction
with electromagnetic radiation are
\begin{eqnarray}\label{9}
\frac{d ~E'}{d~ \tau} &=& E_{i} ' ~-~ E_{o} ' = 0 ~,
\nonumber \\
\frac{d ~\vec{p'}}{d~ \tau} &=& \vec{p}_{i} ' ~-~ \vec{p}_{o} ' ~.
\end{eqnarray}

As for the condition for energy in our Eq. (8), it is equivalent to
Eq. (121) in Kla\v{c}ka (1992), as for the condition for
momentum $\vec{p}_{o} '$ in our Eq. (8), it is a generalization of
Eq. (122) in Kla\v{c}ka (1992).

\section{Stationary frame of reference}
By the term ``stationary frame of reference'' we mean a frame of reference
in which particle moves with a velocity vector $\vec{v} = \vec{v} (t)$.
The physical quantities measured in the stationary frame of reference
will be denoted by unprimed symbols.

Our aim is to derive equation of motion for the particle in the
stationary frame of reference.
We will use the fact that we know
this equation in the proper frame of reference -- see Eqs. (8) and (9).
We have to use Lorentz transformation for the purpose of making
transformation from proper frame of reference to stationary frame
of reference.

If we have a four-vector $A^{\mu} = ( A^{0}, \vec{A} )$, where
$A^{0}$ is its time component and $\vec{A}$ is its spatial component,
generalized special Lorentz transformation yields
\begin{eqnarray}\label{10}
A^{0 '}  &=& \gamma ~ ( A^{0} ~-~ \vec{v} \cdot \vec{A} / c ) ~,
\nonumber \\
\vec{A} ' &=& \vec{A} ~+~ [ ( \gamma ~-~ 1 ) ~ \vec{v} \cdot \vec{A}  /
	      \vec{v} ^{2} ~-~ \gamma ~ A^{0} / c ] ~ \vec{v}  ~.
\end{eqnarray}
The inverse generalized special Lorentz transformation is
\begin{eqnarray}\label{11}
A^{0}  &=& \gamma ~ ( A^{0 '} ~+~ \vec{v} \cdot \vec{A} ' / c ) ~,
\nonumber \\
\vec{A}  &=& \vec{A} ' ~+~ [ ( \gamma ~-~ 1 ) ~ \vec{v} \cdot \vec{A} ' /
	      \vec{v} ^{2} ~+~ \gamma ~ A^{0 '} / c ] ~ \vec{v}  ~.
\end{eqnarray}
The $\gamma$ factor is given by the well-known relation
\begin{equation}\label{12}
\gamma = 1 / \sqrt{1~-~\vec{v} ^{2} / c ^{2} } ~.
\end{equation}

As for four-vectors we immediately introduce four-momentum:
\begin{equation}\label{13}
p^{\mu} = ( p^{0}, \vec{p} ) \equiv ( E / c, \vec{p} ) ~.
\end{equation}

\subsection{Incident radiation}
Applying Eqs. (11) and (13) to quantity $( E_{i} ' / c, \vec{p}_{i} ')$
(four-momentum per unit time -- proper time is a scalar quantity) and
taking into account also Eqs. (8), we can write
\begin{eqnarray}\label{14}
E_{i}  &=& E_{i} ' ~ \gamma ~ ( 1 ~+~ \vec{v} \cdot \hat{\vec{S}_{i}} ' / c ) ~,
\nonumber \\
\vec{p}_{i}  &=& \frac{E_{i} '}{c}  ~ \left \{ \hat{\vec{S}_{i}} ' ~+~
		 \left [ \left ( \gamma ~-~ 1 \right ) ~
		 \vec{v} \cdot \hat{\vec{S}_{i}} '  /
		 \vec{v} ^{2} ~+~ \gamma / c \right ] ~ \vec{v} \right \} ~.
\end{eqnarray}

Using the fact that
$p^{\mu} = ( h ~\nu , h ~\nu ~ \hat{\vec{S}_{i}} )$ for photon,
Lorentz transformation yields
\begin{eqnarray}\label{15}
\nu ' &=& \nu  ~w ~,
\nonumber \\
\hat{\vec{S}_{i}} ' &=& \frac{1}{w}  ~ \left \{ \hat{\vec{S}_{i}} ~+~
		 \left [ \left ( \gamma ~-~ 1 \right ) ~
		 \vec{v} \cdot \hat{\vec{S}_{i}}  /
		 \vec{v} ^{2} ~-~ \gamma / c \right ] ~ \vec{v} \right \} ~,
\end{eqnarray}
where abbreviation
\begin{equation}\label{16}
w \equiv \gamma ~ ( 1 ~-~ \vec{v} \cdot \hat{\vec{S}_{i}} / c )
\end{equation}
is used.

Inserting the second of Eqs. (15) into Eqs. (14), one obtains
\begin{eqnarray}\label{17}
E_{i} &=& ( 1 / w ) ~ E_{i} ' ~,
\nonumber \\
\vec{p}_{i} &=& ( 1 / w ) ~ ( E_{i} ' / c ) ~ \hat{\vec{S}_{i}} ~.
\end{eqnarray}
We have four-vector
$p_{i}^{\mu} = ( E_{i} / c, \vec{p}_{i} ) = ( 1, \hat{\vec{S}_{i}} ) E_{i} / c$
$= ( 1 / w, \hat{\vec{S}_{i}} / w ) w~ E_{i} / c$
$\equiv b_{i}^{\mu} ~ w~ E_{i} / c$.

\subsection{Outgoing radiation}
The situation is analogous to the situation in the preceding subsection.
It is only a little more algebraically complicated, since radiation
may spread out also in directions given by unit vectors
$\hat{\vec{e}_{1}}$, $\hat{\vec{e}_{2}}$. The relations between
$\hat{\vec{e}_{1}} '$ and $\hat{\vec{e}_{1}}$,
$\hat{\vec{e}_{2}} '$ and $\hat{\vec{e}_{2}}$, are analogous to that
presented by the second of Eqs. (15):
\begin{equation}\label{18}
\hat{\vec{e}_{j}} ' = \frac{1}{w_{j}}  ~ \left \{ \hat{\vec{e}_{j}} ~+~
		 \left [ \left ( \gamma ~-~ 1 \right ) ~
		 \vec{v} \cdot \hat{\vec{e}_{j}}  /
		 \vec{v} ^{2} ~-~ \gamma / c \right ] ~ \vec{v} \right \} ~,
		 ~~ j = 1, 2 ~,
\end{equation}
where
\begin{equation}\label{19}
w_{j} \equiv \gamma ~ ( 1 ~-~ \vec{v} \cdot \hat{\vec{e}_{j}} / c )
	      ~, ~~ j = 1, 2 ~.
\end{equation}

Using Eqs. (11) and (13) to quantity $( E_{o} ' / c, \vec{p}_{o} ')$
(four-momentum per unit time -- proper time is a scalar quantity), we
can write
\begin{eqnarray}\label{20}
E_{o}  &=&  \gamma ~ ( E_{o} ' ~+~ \vec{v} \cdot \vec{p}_{o} ' ) ~,
\nonumber \\
\vec{p}_{o}  &=& \vec{p}_{o} ' ~+~
		 \left [ \left ( \gamma ~-~ 1 \right ) ~
		 \vec{v} \cdot \vec{p}_{o} ' /
		 \vec{v} ^{2} ~+~ \gamma ~ \frac{E_{o} '}{c^{2}}
		~\right ] ~ \vec{v}  ~.
\end{eqnarray}
Using also $\vec{p}_{i} '= E_{i} ' ~\hat{\vec{S}_{i}} '/ c$, Eqs. (8), (18),
(20) and the second of Eqs. (15) yield
\begin{eqnarray}\label{21}
E_{o}  &=& Q_{R} ' ~ w ~ E_{i} ~ \gamma  ~+~ ( 1 ~-~ Q_{R} ' ) ~ E_{i} ~+~
\nonumber \\
& &	   w ~ E_{i} ~ ( Q_{1} ' ~+~ Q_{2} ' ) ~ \gamma  ~-~
	   w ~ E_{i} ~ ( Q_{1} ' / w_{1} ~+~ Q_{2} ' / w_{2} )  ~,
\nonumber \\
\vec{p}_{o}  &=& ( 1 ~-~ Q_{R} ' ) ~ \frac{E_{i}}{c} ~\hat{\vec{S}_{i}}  ~+~
		 Q_{R} ' ~ \frac{w~E_{i}}{c^{2}} ~\gamma ~ \vec{v} ~-~
		 \sum_{j=1}^{2} ~ Q_{j} ' ~\frac{w~E_{i}}{c^{2}} ~ \left (
		 c ~ \hat{\vec{e}_{j}} / w_{j}
		 ~-~ \gamma ~ \vec{v}  \right )  ~.
\end{eqnarray}

\subsection{Equation of motion}
In analogy with Eqs. (9), we have for
the changes of energy and momentum of the particle due to the interaction
with electromagnetic radiation
\begin{eqnarray}\label{22}
\frac{d ~E}{d~ \tau} &=& E_{i}	~-~ E_{o} ~,
\nonumber \\
\frac{d ~\vec{p}}{d~ \tau} &=& \vec{p}_{i}  ~-~ \vec{p}_{o}  ~.
\end{eqnarray}
Putting Eqs. (21) into Eqs. (22), using also
$\vec{p}_{i} = ( E_{i} / c ) ~\hat{\vec{S}_{i}}$,
one easily obtains
\begin{eqnarray}\label{23}
\frac{d ~E}{d~ \tau} &=&  Q_{R} ' ~ ( E_{i} ~-~ w ~ E_{i} ~ \gamma ) ~+~
	  \sum_{j=1}^{2} ~Q_{j} ' ~ w ~ E_{i} ~
	  \left ( 1 / w_{j}  ~-~ \gamma \right ) ~,
\nonumber \\
\frac{d ~\vec{p}}{d~ \tau} &=& Q_{R} ' ~ \left \{ \frac{E_{i}}{c}
	~\hat{\vec{S}_{i}}  ~-~ \frac{w~E_{i}}{c^{2}} ~\gamma \vec{v} \right \}
	~+~ \sum_{j=1}^{2} ~ Q_{j} ' ~\frac{w~E_{i}}{c^{2}} ~ \left (
		 c ~ \hat{\vec{e}_{j}} / w_{j}
		 ~-~ \gamma \vec{v}  \right )  ~.
\end{eqnarray}

Equations (23) may be rewritten in terms of four-vectors:
\begin{eqnarray}\label{24}
\frac{d ~p^{\mu}}{d~ \tau} &=&	Q_{R} ' ~ \left ( p_{i}^{\mu} ~-~
	 \frac{w~E_{i}}{c^{2}} ~ u^{\mu}  \right ) ~+~
	 \sum_{j=1}^{2} ~Q_{j} ' ~\frac{w~E_{i}}{c^{2}} ~ \left (
	 c ~ b_{j}^{\mu} ~-~ u^{\mu}  \right ) ~,
\end{eqnarray}
where $p^{\mu}$ is four-vector of the particle of mass $m$
\begin{equation}\label{25}
p^{\mu} = m~ u^{\mu} ~,
\end{equation}
four-vector of the world-velocity of the particle is
\begin{equation}\label{26}
u^{\mu} = ( \gamma ~c, \gamma ~ \vec{v} ) ~.
\end{equation}
We have also found two new four-vectors
\begin{equation}\label{27}
b_{j}^{\mu} = ( 1 / w_{j} , \hat{\vec{e}_{j}} / w_{j} ) ~, ~~ j= 1, 2 ~.
\end{equation}

It can be easily verified that:\\
i) the quantity $w~E_{i}$ is a scalar quantity
-- see first of Eqs. (17); \\
ii) Eq. (24) reduces to
Eq. (7) and to the first of Eqs. (9) for the case of proper inertial
frame of reference of the particle; \\
iii) Eq. (24) yields $d ~m / d~ \tau =$ 0.

\section{Conclusion}
We have derived equation of motion for real dust particle under the action
of electromagnetic radiation. It is supposed that equation of motion is
represented by Eqs. (8) and (9) in the proper frame of reference of the
particle. The final covariant form is represented by
Eq. (24), or, using the relations $E_{i} = w ~ S ~A'$ (see Eq. (38) in
Kla\v{c}ka 1992) and
$\vec{p}_{i} = ( E_{i} / c ) ~\hat{\vec{S}_{i}}$
\begin{eqnarray}\label{28}
\frac{d ~p^{\mu}}{d~ \tau} &=&	\frac{w^{2} ~ S ~A'}{c^{2}} ~ \left \{
	 Q_{R} ' ~ \left ( c~ b_{i}^{\mu} ~-~u^{\mu}  \right ) ~+~
	 \sum_{j=1}^{2} ~Q_{j} ' ~ \left (
	 c ~ b_{j}^{\mu} ~-~ u^{\mu}  \right ) \right \}
\end{eqnarray}
(four-vector $b_{i}^{\mu}$ is defined below Eq. (17)).

Within the accuracy to the first order in $\vec{v} / c$, Eq. (28) yields
\begin{eqnarray}\label{29}
\frac{d~ \vec{v}}{d ~t} &=&  \frac{S ~A'}{m~c} ~ \left \{ Q_{R} ' ~ \left [
		 \left ( 1~-~ \vec{v} \cdot \hat{\vec{S}_{i}} / c \right ) ~
		 \hat{\vec{S}_{i}} ~-~ \vec{v} / c \right ] ~+~ \right .
\nonumber \\
& &  \left .  \sum_{j=1}^{2} ~Q_{j} ' ~\left [  \left ( 1~-~ 2~
	      \vec{v} \cdot \hat{\vec{S}_{i}} / c ~+~
	      \vec{v} \cdot \hat{\vec{e}_{j}} / c \right ) ~ \hat{\vec{e}_{j}}
	      ~-~ \vec{v} / c \right ] \right \} ~.
\end{eqnarray}
As for practical applications, the terms $v/c$ standing at $Q_{j}'$
are negligible for majority of real particles.
(We want to stress that values
of $Q'-$coefficients depend on particle's orientation with respect to the
incident radiation -- their values are time dependent.)

\acknowledgements{}
The author wants to thank to H. Kimura, H. Okamoto and T. Mukai
for their remark that for the purpose of light scattering theory
the transformation $p' ( \vartheta ', \varphi ')$ $\rightarrow$
$C_{sca} ^{'-1} ~ dC'_{sca} / d \Omega '$ is required, where
$C'_{sca}$ denotes scattering cross section and
$dC'_{sca} / d \Omega '$ is the differential scattering cross section.
(June 2001; see, e. g., H. Kimura and I. Mann: 1998, Radiation pressure
cross section for fluffy agregates, {\it J. Quant. Spetrosc. Radiat. Transfer}
{\bf 60/3}, 425-438)

\end{document}